\documentclass[conference]{IEEEtran}
\IEEEoverridecommandlockouts
\usepackage{hyperref}
\usepackage{cite}
\usepackage{amsmath,amssymb,amsfonts}
\usepackage{algorithmic}
\usepackage{graphicx}
\usepackage{textcomp}
\usepackage{xcolor}
\usepackage{float}
\usepackage{subcaption} %  for subfigures environments 
\def\BibTeX{{\rm B\kern-.05em{\sc i\kern-.025em b}\kern-.08em
    T\kern-.1667em\lower.7ex\hbox{E}\kern-.125emX}}

\usepackage{pifont}
\usepackage{colortbl}
\usepackage{braket}
\usepackage{comment}
\usepackage{tikz}
\usetikzlibrary{quantikz2}

\newcommand{\CellWithForceBreak}[2][c]{
\begin{tabular}[#1]{@{}c@{}}#2\end{tabular}}

\usepackage{balance}
\usepackage{xcolor}
\usepackage[ruled,vlined,noend]{algorithm2e}

\SetCommentSty{mycommfont}
\SetAlFnt{\small}
\SetAlCapFnt{\small}
\SetAlCapNameFnt{\small}

\usepackage{soul}
\def\BibTeX{{\rm B\kern-.05em{\sc i\kern-.025em b}\kern-.08em
    T\kern-.1667em\lower.7ex\hbox{E}\kern-.125emX}}

\begin{document}

%\title{Stealthy Conditional Trojans in Quantum Circuits}
\title{Quantum Trojan Insertion: Controlled Activation for Covert Circuit Manipulation}

\author{
    \IEEEauthorblockN{Jayden John, Lakshman Golla, Qian Wang}  
    \IEEEauthorblockA{Department of Electrical Engineering, University of California, Merced, CA, USA}  
    \IEEEauthorblockA{\{jjohn92, lgolla, qianwang\}@ucmerced.edu}  
}

% \author{
%     \IEEEauthorblockN{Jayden John\IEEEauthorrefmark{}, Lakshman Golla\IEEEauthorrefmark{},  Qian Wang\IEEEauthorrefmark{}}
%     \IEEEauthorblockA{\IEEEauthorrefmark{}Department of Electrical Engineering, University of California, Merced, CA, USA}
%     \IEEEauthorblockA{{jjohn92, lgolla, qianwang}@ucmerced.edu}
% }

\maketitle

%\title{Quantum Logic Locking (QLL): Safeguarding Intellectual Property in Quantum Circuit Designs}

%\email{webmaster@marysville-ohio.com}
%\affiliation{%
%  \institution{Institute for Clarity in Documentation}
%  \country{USA}
%}

\begin{abstract}
Quantum computing has demonstrated superior efficiency compared to classical computing. Quantum circuits are essential for implementing functions and achieving correct computational outcomes. Quantum circuit compilers, which translate high-level quantum operations into hardware-specific gates while optimizing performance, serve as the interface between the quantum software stack and physical quantum machines. However, untrusted compilers can introduce malicious hardware Trojans into quantum circuits, altering their functionality and leading to incorrect results. In the world of classical computing, effective hardware Trojans are a critical threat to integrated circuits. This process often involves stealthily inserting conditional logic gates that activate under specific input conditions. In this paper, we propose a novel advanced quantum Trojan that is controllable, allowing it to be activated or deactivated under different circumstances. These Trojans remain dormant until triggered by predefined input conditions, making detection challenging. Through a series of benchmark experiments, we demonstrate the feasibility of this method by evaluating the effectiveness of embedding controlled trojans in quantum circuits and measuring their impact on circuit performance and security.

\end{abstract}

\maketitle

\section{Introduction}
Quantum computers have garnered significant attention in recent years, driven by the availability of quantum computing services provided by industry leaders such as IBM Quantum \cite{chow2021ibm}, Amazon Braket \cite{gonzalez2021cloud}, and Microsoft Azure \cite{prateek2023quantum}. Unlike classical computers, which are typically accessible and physically close to users, quantum computers are placed in specialized facilities requiring stringent cooling systems and are accessible only through remote interfaces. Additionally, quantum computers cannot directly process user code or programs. Instead, a translation layer, known as a quantum compiler or transpiler, is required to convert high-level quantum programs into instructions that are compatible with the quantum hardware. This compilation process involves transforming the original quantum circuit into an executable format by replacing gates with basis ones, inserting swap gates to accommodate physical qubit connectivity, and incorporating optimizations to enhance the performance of the compiled circuits.

However, the compilation process may introduce vulnerabilities, as malicious compilers have the potential to exploit and manipulate quantum circuit designs for unintended purposes. These threats include the insertion of Trojans \cite{das2023trojannet,roy2024hardware} and the counterfeiting of quantum designs \cite{yang2024multi}. Many quantum compilers are provided by third-party vendors, which cannot be fully trusted by quantum computer providers or their users. Malicious activities embedded within quantum programs could significantly impact the actual execution of the circuits on quantum machines.

Recent attacks on quantum systems have focused primarily on pulse-level side-channel attacks \cite{xu2023exploration,trochatos2024dynamic} and intellectual property (IP) protection of quantum circuits \cite{ghosh2023primer,das2024secure}. Previous studies have explored quantum Trojans but have largely been limited to basic forms, such as the insertion of X-gates or swap gates without triggers \cite{upadhyay2024stealthy,das2024Trojan}. These basic Trojans are easily detectable and removable through compiler optimization functions or machine learning-based detection techniques \cite{das2023trojannet,roy2024hardware}. In classical circuit design, stealthy hardware trojans are often embedded using conditional logic, making them dormant and indistinguishable from normal circuits until triggered by specific conditions. This work aims to address the gap in stealthy Trojan designs for quantum circuits by introducing a new class of quantum Trojans. These Trojans leverage the concept of remaining hidden from detection through controlled gate operations and are triggered only under predefined input conditions.

In this paper, we analyze the properties of quantum circuits and the effects of the transpilation process to propose a novel type of quantum Trojan. This enhances stealth and resilience against existing detection and optimization techniques, significantly advancing the state of quantum hardware Trojan research.
The paper has the contribution as follows.

\begin{itemize}
   \item We introduce controlled Trojan insertion in quantum circuits, where the activation signals trigger the Trojan only under specific conditions. This approach enhances stealth and resilience against existing detection and optimization techniques.
   \item We introduce a strategic process for selecting optimal positions to insert Trojan gates while preserving the circuit's correct functionality when deactivated. More importantly, this ensures that the inserted circuit maintains the same depth as the original, avoiding any additional depth overhead.
   \item Experimental results on the \textit{RevLib} benchmark set demonstrate that the injection of Trojan gates achieves a 0\% depth increase and a total variation distance approximately 90\% from the original circuits. 
   
\end{itemize}

\section{Background \& Related Work}

\subsection{Hardware Trojans}
\label{ssec:hardware Trojans}
Introducing hardware Trojans into silicon chips is a significant security concern in the semiconductor supply chain. Most IC design companies rely on third-party foundries for chip fabrication, making the design process vulnerable to malicious modifications during manufacturing. 
%Hardware Trojans, often inserted during this stage, can alter the chip's functionality, leak sensitive information, or compromise its reliability. The primary security goal in countering hardware Trojans is to prevent unauthorized modifications or malicious insertions during the manufacturing process.
Hardware Trojan attacks can take various forms depending on the stage of the design and fabrication process. In traditional silicon chip manufacturing, the threat often arises during the fabrication phase, where an untrusted foundry can insert malicious circuitry. Trojans can be classified into combinational and sequential types, with activation mechanisms ranging from rare input conditions to environmental triggers. 
%For example, a simple combinational Trojan might add logic gates to manipulate data, while more sophisticated Trojans might integrate state-based designs that remain dormant until specific conditions are met.

Researchers have explored a variety of techniques for embedding Trojans, ranging from combinational logic modifications \cite{karri2008trustworthy} to sequential state-based designs that activate only under specific conditions, such as rare input patterns or environmental triggers \cite{bhunia2014hardware}.
Despite these efforts, stealthy Trojans—those designed to evade detection by mimicking legitimate functionality or remaining dormant until activation—continue to pose significant challenges \cite{forte2017comprehensive}.

To counter these threats, detection methods have been developed, including side-channel analysis, which monitors power, delay, or electromagnetic emissions \cite{jin2008power, yang2016improved}, and logic testing approaches that identify anomalous behaviors in circuit functionality \cite{chakraborty2009testing}. Advanced techniques, such as machine learning-based detection \cite{yasin2017machine} and runtime monitoring \cite{zhang2014runtime}, have further improved Trojan detection accuracy.

\subsection{Quantum Circuits} \label{ssec:quantum_circuits}

A quantum circuit is composed of a series of quantum gates (functional units) and that manipulate the states of the quantum basic state-- (a.k.a. qubit). A qubit has two basis states, denoted by the bracket notation as $\ket 0$ and $\ket 1$. 
The state $\ket \psi$  can be written as a linear function of the basis state as  $\ket \psi = \alpha \ket 0 + \beta \ket 1 = [\alpha, \beta]^T$. 

Similar to classical logic gates, the fundamental actions at the logic level in quantum computing are performed by quantum gates. These gates carry out unitary operations, operator as $U$, which transform the states of input qubit. Quantum algorithms are composed of sequences of these quantum gates, designed to evolve input qubits into desired quantum states.  A quantum gate $U$ operating on a qubit $\ket \psi$ can be written down as $\ket \psi \rightarrow U \ket \psi$. 

Single qubit gates operate on a single qubit and are analogous to the elementary logic gates in classical computing, such as the NOT gate. These gates manipulate a qubit by rotating its state vector on the Bloch sphere, a geometric representation of its state. The most common single qubit gates include the Pauli gates (X, Y, Z), the Hadamard gate (H), and phase shift gates (S, T).

Multi-qubit gates are essential in quantum computing because they enable interactions between qubits to facilitate complex operations in quantum algorithms. For example, the Controlled-NOT (CNOT) gate operates on two qubits: a control qubit and a target qubit. The control qubit would determine whether the operation is applied to the target qubit. The target qubit's state is flipped by the control qubit in the $\ket 1$ state. 
Many quantum algorithms rely heavily on the CNOT gate for their implementation, especially in constructing entangled states and performing conditional operations.

Complex quantum gates can be decomposed into fundamental single- and two-qubit gates, with the choice of basic gate types depending on the specific quantum architecture. For example, IBM quantum computers use 
ID, RZ, SX and X gates as their single-qubit basis gates. For multi-qubit operations, IBM quantum machines typically use the CNOT  as the basis two-qubit gate. These gates form the foundational set for compiling and executing quantum circuits on IBM's quantum hardware.

\begin{figure}[htb]
\captionsetup{font=small} 
    \centering
    \[
\begin{quantikz}
\lstick{\( q_0 \)} & \gate{H}     & \ctrl{1} & \qw      & \meter{} \\
\lstick{\( q_1 \)} & \qw & \targ    & \qw   &\qw   & \meter{}
\end{quantikz}
\]
    \caption{An example of a quantum circuit}
    \label{fig:quantum_circuit}
\end{figure}
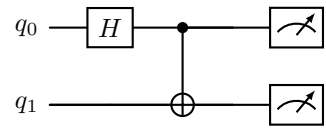

A quantum circuit is a sequence of quantum gates arranged to perform a computation on one or more qubits. It typically starts with qubits initialized in a known state (e.g., $\ket{0}$), followed by the application of single and multiple qubit gates. The circuit can be visualized as a timeline with wires representing qubits and gate operations along them, shown as in Figure \ref{fig:quantum_circuit}.

\subsection{Quantum Circuit Compilation and Trojans}
Quantum circuit compilation transforms a circuit representing a quantum algorithm into a hardware-executable format, adapting it to the quantum hardware topology, connectivity, and error characteristics of the target quantum hardware. This process is analogous to compiling classical programs but must address unique quantum constraints such as superposition, entanglement, and limited qubit coherence times. Efficient compilation ensures that quantum algorithms are executed with minimal resource overhead, reduced error rates, and optimal performance. Optimization techniques in compilation focus on reducing gate count, circuit depth, and qubit interactions, thereby enhancing fidelity and performance on noisy intermediate-scale quantum (NISQ) devices.

However, the open-source nature of many quantum compilers introduces potential security vulnerabilities. These compilers, often developed and maintained by untrusted third-party entities, can become malicious activities. Adversaries may exploit this by inserting quantum Trojans during the compilation process, which can disrupt circuit functionality and degrade performance. Such attacks can result in incorrect outputs or even circuit failures if the hardware's capacity is exceeded.

\begin{figure*}[tb]
\captionsetup{font=small} 
    \centering
    \includegraphics[width=0.85 \textwidth]{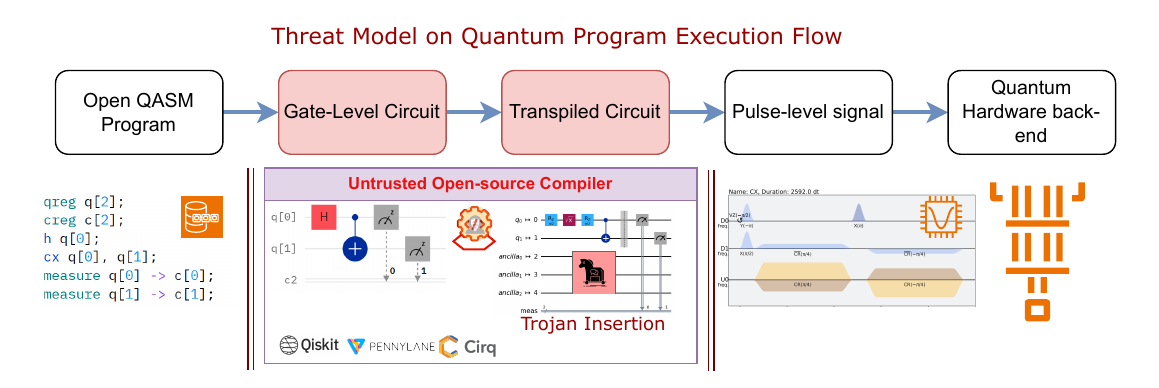}
    \caption{Illustration of the untrusted compiler threat model. During the compile process, the malicious Trojans are inserted.}
    \label{fig:threat_model}
\end{figure*}

\subsection{Related Works}
Quantum circuits have emerged as a critical area of security research due to their importance in the era of quantum computing. Several studies have investigated the challenges and vulnerabilities associated with quantum circuits \cite{ghosh2023primer}. One significant area of research focuses on IP protection for quantum circuits, it mainly addresses concerns similar to those in classical integrated circuit (IC) design, such as IP theft \cite{aboy2022mapping}. Related work has also explored techniques such as quantum logic locking and quantum circuit obfuscation to protect quantum circuits against IP-related attacks \cite{das2023randomized, suresh2021short, topaloglu2023quantum}. These techniques aim to secure the integrity and confidentiality of quantum designs.
Other studies have investigated insertion-based attacks designed to disrupt the quantum computing process. For instance, prior work \cite{roy2024hardware} introduced the concept of Trojan insertion targeting quantum circuits and proposed a CNN-based detection method to identify these malicious modifications \cite{das2023trojannet}. Another study investigated the introduction of adversarial SWAP gates to amplify the computational burden during quantum compilation. \cite{upadhyay2024stealthy}, effectively degrading performance. These efforts have laid a crucial foundation for understanding insertion attacks and their significant impact on quantum circuits.
In this paper, we build upon prior research by introducing a novel approach to controlled Trojan insertion in quantum circuits. Unlike earlier methods involving single-qubit Trojans, our design enables the Trojans to remain dormant until triggered by specific control signal conditions, akin to the behavior of hardware Trojans in classical IC design. This conditional activation significantly improves concealment, making the Trojan harder to detect and more resilient against removal during optimization. For instance, simple gate-based Trojans, such as redundant X-gates, are often eliminated during the compilation process. In contrast, our approach embeds conditional logic directly into the circuit, ensuring the Trojan remains functional and hidden until its activation criteria are met.

\section{Threat Models}
Running quantum programs on real quantum hardware is often prohibitively expensive and time-consuming, demanding substantial resources for testing and evaluation. Moreover, achieving accuracy in the realized functions is crucial, particularly given the inherent noise and error-prone nature of quantum computing systems. Given the open-source nature of quantum compilers, these tools may pose potential security risks as they are often developed or maintained by untrusted third-party entities. Many quantum programs rely on these compilers for tasks such as optimization or error correction \cite{smith2020open, salm2021automating}, introducing vulnerabilities that adversaries can exploit. Specifically, malicious actors could tamper with the compilation process by inserting or modifying quantum gates. These insertions, referred to as quantum Trojans, can disrupt circuit functionality or increase computational overhead, requiring additional trials on quantum hardware. 

This paper assumes a threat model where an untrusted third-party compiler is on a remote server and with an adversary's control over the compilation process. The adversaries with prior experience with quantum circuits can strategically select and place Trojan gates to maximize disruption. The adversaries' capabilities include access to the quantum program, the ability to perform resource analysis, and limited knowledge of the quantum function. Quantum circuits typically undergo transpilation to adapt to the topology and constraints of the target quantum hardware. This process often results in a transformed circuit layout. In our threat model, we assume the adversary has access to both the original circuit and the transpiled version. The adversary can choose to insert malicious gates into the original circuit, the compiled circuit, or both. These assumptions align with previous research on quantum circuit security \cite{roy2024hardware,upadhyay2024stealthy}.

\section{Quantum Trojan and Analysis}

\begin{figure}[htb]
    \centering
    \captionsetup{font=small} 
    \includegraphics[width=0.9\linewidth, trim={0.5cm, 0.5cm, 2.5cm, 1.5cm}, clip]{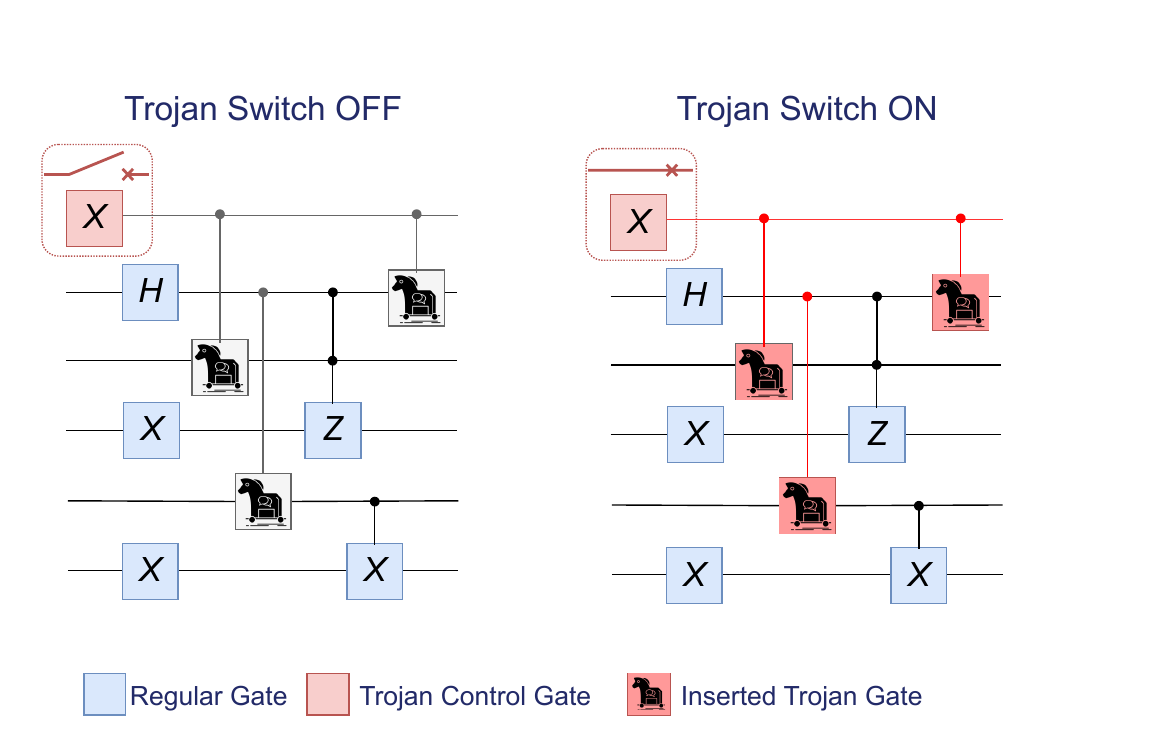}
    \caption{Example of Trojan insertion. When the switch is off (left), the circuit operates normally, maintaining its intended functionality. However, when the switch is on (right), the Trojan gates become active, altering the circuit's behavior. }
    \label{fig:trojan}
\end{figure}

The design of the quantum circuit is sent to a third-party untrusted compiler for compilation, where malicious attackers may insert Trojan gates. Our designed Trojan can be selected to be activated using a trojan control gate. When the control gate is switched off, the circuit executes its intended quantum computations through standard gate operations, shown in Figure \ref{fig:trojan}. The arrangement ensures that during normal operation, the circuit maintains its expected functionality, making the presence of the Trojan difficult to detect through standard verification procedures. When the Trojan is activated, the additional gates introduce controlled modifications to the quantum computations.

\subsection{Trojan Insertion Process}
\label{ssec:qll_method}

The quantum trojan insertion methodology employs a sophisticated two-phase approach for integrating controlled logic into quantum circuits while maintaining the original function. This process systematically identifies and utilizes \textit{empty positions} in the circuit to implement disguised gate insertions.

\begin{figure}[htb]
\captionsetup{font=small} 
    \centering
    \[
\begin{quantikz}[slice all]
\lstick{$q_0$} & \gate{X} & \ctrl{1} & \ctrl{2} & \targ{} & \ctrl{1} & \gate[style={dashed, fill=pink!20, draw=red}] {\emptyset} & \ctrl{2} & \qw \\
\lstick{$q_1$} & \gate[style={dashed, fill=pink!20, draw=red}] {\emptyset} & \targ{} & \ctrl{1} & \ctrl{-1} & \targ{} & \ctrl{2} & \ctrl{1} & \qw \\
\lstick{$q_2$} & \gate{X} & \gate[style={dashed, fill=pink!20, draw=red}] {\emptyset} & \targ{} & \ctrl{-2} & \ctrl{-1} & \ctrl{1} & \targ{} & \qw \\
\lstick{$q_3$} & \gate[style={dashed, fill=pink!20, draw=red}] {\emptyset} & \ctrl{-2} & \ctrl{-1} & \ctrl{-3} & \ctrl{-2} & \targ{} & \ctrl{-1} & \meter{}
\end{quantikz}
\]
    \caption{The dashed pink boxes labeled `$\emptyset$` indicate empty slots (no operation) on the respective qubit in that column.}
    \label{fig:quantum_circuit2}
\end{figure}
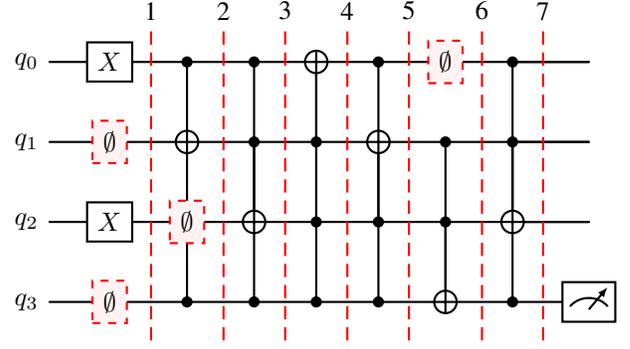

To effectively avoid significant increases in circuit depth and gate count, we employ an algorithm to identify empty slots in the original circuits at each layer. Trojans are then inserted into these empty slots, as detailed in Algorithm 1. To identify the empty positions, we convert the target quantum circuit into a directed acyclic graph (DAG) representation, which provides a structured framework for analyzing the circuit's dependencies and available insertion points. This conversion facilitates granular analysis of quantum operations across distinct temporal layers. Within each layer, the algorithm enumerates the qubits to dynamically identify the set of utilized qubits. 
Through set-theoretic complementation, where $E = Q \backslash S$ ($Q$ representing the complete qubit set and $S$ the utilized qubits), the algorithm derives empty positions set as $E$. This systematic identification ensures that potential insertion points do not interfere with existing quantum operations to preserve the circuit's original computational structure. An example illustrated the identification of layers of the circuit as well as the empty spots shown in Figure \ref{fig:quantum_circuit2}.

The second phase introduces a significant refinement in the gate insertion strategy. At the circuit's control qubit (index 0), the algorithm implements the switch gate (X-gate) at a predetermined control position. This gate serves as the primary trigger mechanism for the Trojan's activation. For subsequent columns, we strategically place controlled-NOT (CX) gates at the predetermined control positions, while the target qubit is randomly selected from the available empty positions.

The insertion process maintains strict adherence to the specified gate limits. The available qubit positions are continuously updated through intersection operations with identified empty positions, and selected positions are removed from the available pool after gate insertion. This dynamic resource management ensures the integrity of the circuit while successfully integrating controlled modifications. The approach demonstrates enhanced control over Trojan activation through the explicit designation of control positions and the systematic placement of corresponding quantum gates. This architectural modification enables more precise trigger conditions while maintaining the stealth characteristics essential for hardware Trojan implementation in quantum circuits.

\begin{algorithm}[tb]
\DontPrintSemicolon
\SetAlgoLined
\SetNoFillComment
\LinesNotNumbered 
\caption{Random gate insertion into empty positions}\label{alg:empty_positions}
\KwData{\textbf{C}: quantum circuit, \textbf{empty\_pos}: List to store empty positions}
\KwIn{ total\_qubits, gate\_limit, control\_pos}
\tcc{Step1:Get empty positions}
Convert circuit to DAG representation and Extract layers\;

\For{each layer in  \textbf{layers}}{
    Get operations in the current layer\;
    Initialize empty set for used qubits\;
    
    \For{each operation in layer}{
        used\_qubits $\gets$ GetQubitIndices(operation)\;
    }
    empty\_positions $\gets$ sorted(list(all\_qubits $\setminus$ used\_qubits))
}
\tcc{Step2: randomly insertion into circuits}

\ForEach{column $\in$ quantum circuits}{
    \If{added\_gates $>$ gate\_limit}{
    Break\;
    }
    available\_qubits $\leftarrow$ available\_qubits $\cap$ empty\_positions\;
    %Sort(available\_qubits)\;
    \If{column.index==0}{
        %find(q1, q2) $\in$ available\_qubits\;
        AddXGate(circuit, control\_pos) \tcp*[l]{insert control gate}
    }
    \Else{\tcp*[l]{insert CX gate}
        random\_pos $\leftarrow$ RandomChoice(available\_qubits)\;
        AddCXGate(circuit, control\_pos, random\_pos)\;
        available\_qubits $\leftarrow$ available\_qubits $\setminus\{$random\_pos$\}$\;
        added\_gates $\leftarrow$ added\_gates $+ 1$\;
        }
    }
\KwRet{\textbf{C}: quantum circuit}\;

\end{algorithm}

\subsection{Improvements to Prior Approach}

Several studies have discussed the vulnerabilities that hardware Trojans bring to quantum circuits and provided considerable information on how to detect and mitigate them. As an example, \cite{das2024Trojan} investigated Trojan attacks on variational quantum circuits, focusing on the Quantum Approximate Optimization Algorithm (QAOA). Their work emphasized the vulnerability of QAOA circuits to the insertion of Trojan gates and proposed a CNN-based detection system that achieved high accuracy for such an attack detection. However, their analysis was restricted to static Trojan insertions and did not extend to exploring the implications of more sophisticated conditional activation of such Trojans. Along similar lines, \cite{roy2024hardware} investigated the impact of single-gate Trojans, including Hadamard and NOT gates, inserted at different locations in quantum circuits. Their results indicated that such Trojans can degrade circuit functionality, especially in noisy environments. While their work provided valuable insights into the outcome of single-gate insertions; it did not consider more sophisticated Trojan designs, such as 2-qubit gates and those that are triggered only under certain conditions.

Another closely related approach is TrojanNet \cite{das2023trojannet}, a machine learning-based framework for detecting Trojan-inserted circuits. The approach focused on the QAOA algorithm and achieved very good detection accuracy. Despite the success in identifying Trojan attacks, the study focused only on finding static modifications of gates. Their detection methods do not account for dynamically triggered Trojans, which are activated based on specific, well-defined input patterns.

Our work deviates from the previous work in terms of presenting a new strategy to embed conditional logic gates into quantum circuits for designing Trojans. Such a Trojan would be conditionally activated while remaining latent under normal operating conditions and activating when predefined input conditions are met. This renders the Trojan far more difficult to detect compared to those that are designed statically. We perform benchmark experiments to test the performance and security implications of these conditional Trojans, demonstrating their feasibility and challenges for existing detection methods. Our findings emphasize the necessity of new detection strategies tailored to address the unique characteristics of conditional hardware Trojans in quantum computing systems.
%\subsection{Security Metrics}
 
%The security metrics for classical logic locking techniques are generally related to the resiliency to known attacks that are discussed in Section \ref{ssec:logic_locking}. However, none of the attacks on classical logic locking would apply to QLL. 
%Here, we propose a scheme to evaluate how effectively a quantum circuit is locked by key values. % The evaluation process is illustrated in the figure. 
%Suppose we have a quantum circuit with an additional qubit key input. This additional key-qubit provides multiple positions where key values can be inserted. The $x$ is defined as the input of the circuit, and the output $y$ is measured as usual at the end of the circuit. In our evaluation, we first select different key inputs $k$, then traverse all possible inputs $x$, comparing the outputs under the same conditions with the correct key value, denoted as $k_{correct}$. This comparison gives us the guessing rate for each input. Finally, we average the guessing rates across all input sets to determine the overall guessing rate.

% \begin{equation}
%      guessRate = \frac{\sum_k ^ {N_k} \sum_{i} ^ {N_x} |y_{k,x_i} - y_{k_{correct},x_i}|} {N * N_k *N_x}
% \end{equation}

\section{Experiments Evaluation}
%\hl{Lakshman}\\
\subsection{Experimental Setup}

We conducted our experiments using the IBM Qiskit framework to compile and simulate quantum circuits. We utilized benchmark circuits from the \textit{RevLib} benchmarks \cite{wille2008revlib} for our experiments, which has been widely utilized in prior work on quantum circuit compilation. These benchmark circuits encompass a diverse set of gate operations, with the number of gates ranging from 4 to 30 and qubit sizes varying across 4, 5, 7, 10, and 12 qubits. 
%The circuits were constructed using the \textit{QuantumRegister} and \textit{QuantumCircuit} modules.
The benchmark circuits have both 1-bit output and multi-bit outputs to simulate the function locking with different output cases.
To recreate proper simulation conditions, we employed the \textit{FakeValencia} backend from Qiskit \cite{qiskit2024}, which incorporates the noise model of the actual \textit{ibmq-valencia} device. All simulations were performed with 1,000 shots to generate statistically significant results. Both the original and Trojan-inserted circuits were simulated using the same backend, ensuring that any observed differences can be attributed to the Trojan mechanism rather than variations in the simulation environment.

As outlined in Section \ref{ssec:qll_method}, we insert the controlled Trojan, comprising a specific set of CX gates and a switch gate, at the beginning of the original circuit. The empty positions for insertion are strategically selected based on the operations present in the benchmark circuits. This tailored approach ensures that the Trojan is embedded in unused empty slots of the circuit's DAG, minimizing hardware overhead and enhancing the overall stealthiness and effectiveness of the modified quantum circuit.

\subsection{Metrics for Evaluation}

\textit{Total Variation Distance (TVD)} is a metric used to measure the distance between two probability distributions. This metric is particularly suitable for quantum circuit measurement because the output is made up of probabilistic distributions. For example, the output of a 1-bit circuit simulation with noise can be represented as a distribution, such as \{``0'': 95, ``1'': 5\}, based on 100 shots. In this context, TVD measures the discrepancy between the output distributions of the correct (original) circuit and the modified circuit. 
It is calculated as the sum of absolute differences between the counts of each outcome in the two distributions, normalized by the total number of shots. The formula for TVD is:
\begin{equation}
    TVD = \frac{\sum_{i=0}^{2^b-1} |y_{i,orig} - y_{i,alter}|}{2N}
\end{equation}
Where $N$ represents the total number of shots in this run, $b$ represents the number of output qubits, resulting in 
$2^b$ possible output types. $y_{i,alter}$ and $y_{i,orig}$ represent the count of value $i$ in the altered and original quantum circuits respectively. 

This discrepancy in TVD value highlights the effectiveness of the inserted Trojans. An effective Trojan will significantly disrupt the functionality of the original circuit, resulting in more bit flips from the original output distribution.

\begin{table*}[h]
\captionsetup{font=small} 
\centering
\begin{tabular}{|l|c|c|c|c|c|c|c|c|c|c|}
\hline
\textbf{Circuit} & \textbf{Depth} & \textbf{\CellWithForceBreak{Depth \\ Obfuscated}} & 
% \textbf{\CellWithForceBreak{Depth \\ Restored }} &  
\textbf{\CellWithForceBreak{Gate \\ Count}} & \textbf{\CellWithForceBreak{Gate \\ Obfuscated}} & 
\textbf{\CellWithForceBreak{Gate \\ difference}} &
\textbf{Accuracy}  & \textbf{\CellWithForceBreak{Accuracy \\ Deactivated}} & \textbf{\CellWithForceBreak{Accuracy \\ change(\%)}} \\ \hline
mini\_ALU & 8  & 8  & 7 & 9 & 2 & 0.983 & 0.943 & -4\% \\ \hline
4mod5    & 6 &6  & 7  & 11 & 4 & 0.937  & 0.994  & +5.7\% \\ \hline
1-bit adder & 5 & 5 & 5  & 7 & 2 & 0.959 & 0.925 & -3.4\% \\ \hline
4gt11    & 13 &13 & 13 & 15 & 2 & 0.986 & 0.983 & -0.30\% \\ \hline
4gt13    & 4  & 4 & 4  & 6 & 2 & 0.948 & 1.00 & +5.2\% \\ \hline
rd53     & 16 & 16 & 16 & 22 & 6 & 0.941 & 0.998  & +5.7\% \\ \hline
rd73     & 13 & 13 & 20 & 25 & 5 & 0.991 & 0.994 & +0.3\% \\ \hline
rd84     & 15 & 15 & 28 & 34 & 6 & 0.993 & 0.992 & -0.1\% \\ \hline
ALU     & 7 & 7 & 7 & 9 & 2 & 0.919 & 0.953 & +3.4\% \\ \hline
sym6     & 13 & 13 & 22 & 27 & 5 & 0.993 & 0.992 & -0.1\% \\ \hline
\end{tabular}
\caption{Comparison of circuit parameters: depth, count, accuracy, and fidelity change before and after alterations, data shown here are the averages of 20 iterations. The original circuit is from the \textit{RevLib}.}
 \vspace{-5mm}
\label{tab:circuit_parameters}
\end{table*}

\subsection{Result Analysis}
\begin{figure}[!htp]
    \centering
    \captionsetup{font=small} 
    \includegraphics[width=0.5\textwidth]{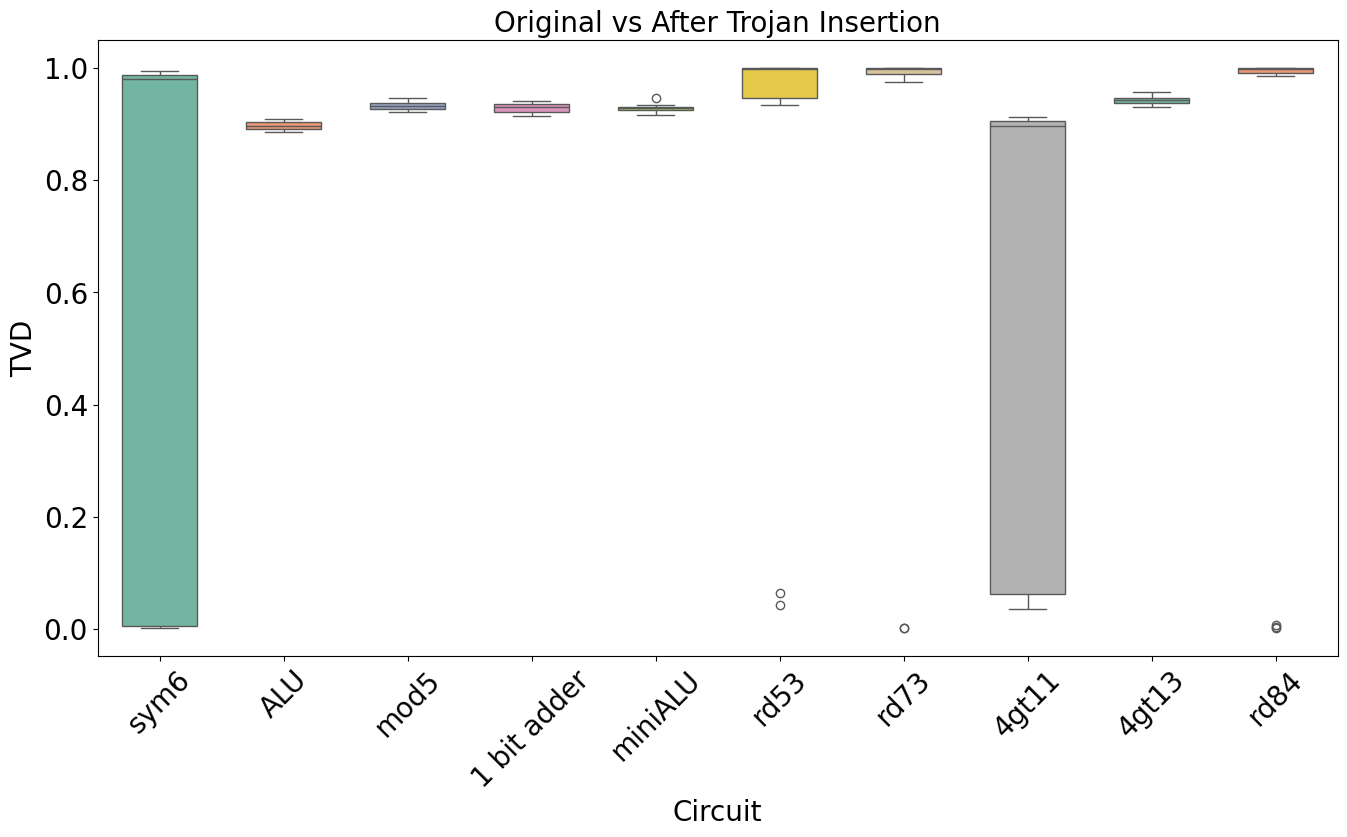}
    \caption{Distribution of Total Variation Distance (TVD) of benchmark circuits: TVD of obfuscated circuit and restored circuit are calculated and shown respectively. Selected circuits are simulated using Qiskit and the FakeValencia backend, incorporating noise into the simulation}
    \label{fig:result_vd}
\end{figure}

This section presents our experimental results from the \textit{RevLib} benchmarks simulated using the Qiskit backend. These results encompass both 1-bit and multi-bit quantum circuits.
Figure \ref{fig:result_vd} illustrates a comparison of TVD values across different circuits after trojan insertion.  TVD is calculated as the variation distance with the theoretical output. For example, in the case of a 1-bit Adder with 100 shots, we use the result as ``0'': 100, ``1'': 0  \} as the reference to compare the relative distance. 

The Total Variation Distance (TVD) values observed after Trojan insertion showed significant variations, indicating notable functional alterations caused by the presence of the Trojan circuit. For most of the circuits, such as rd84, rd73, and rd53, the TVD values approach 1, indicating significant changes in the output distribution. This occurs because these circuits produce multi-bit outputs and are relatively large and deep, providing ample opportunities for inserting random gates. More random gate insertion results in more flips in the output. In contrast, smaller circuits with 1-bit output, such as sym6 and  4gt11, have less significant changes in TVD value. This is because these circuits provide limited space for inserting trojan gates compared to more complex circuits. 
Overall, Figure \ref{fig:result_vd} shows that the altered circuits differ significantly from the original circuits in TVD values. The consistent TVD values across the benchmarks indicate that our insertion method uniformly impacts all circuits.

\subsection{Cost and Overhead Analysis}

The results on gate counts and circuit depth across various circuits are presented in Table \ref{tab:circuit_parameters}. To minimize disruption to the circuit’s original structure, we implemented a selection algorithm that strategically places Trojan gates exclusively in unused or empty slots within the circuit. This approach ensures that the circuit depth remains unchanged after Trojan insertion, preserving its functional timing and critical path integrity. 
The number of gates inserted across the circuits ranges from 2 to 6, resulting in an average 20\% increase in the total gate count. 
For larger and deeper circuits, the percentage increase in gate count diminishes due to the higher baseline number of gates in these circuits.
This approach ensures that while the inserted Trojans add complexity to the circuits, they do not significantly impact the overall computational resources required for larger quantum circuits. As a result, we maintain efficiency and scalability in Trojan insertion, even when dealing with complex and extensive quantum circuits.

\section{Conclusion}

In this paper, we present a novel method for inserting controlled Trojans into quantum circuits. Unlike existing Trojan insertion techniques, our approach activates the Trojans only under specific conditions, enhancing their stealthiness. Experimental results across various quantum circuits demonstrate that our method introduces minimal overhead, with circuit depth remaining unchanged and only a 20\% increase in gate count. This highlights the significant threat posed to the security of quantum circuits by such stealthy Trojan designs.
%\balance

%\bibliographystyle{ACM-Reference-Format}

\bibliographystyle{IEEEtran.bst}
\bibliography{qtrojan}
%\bibliography{reference}
\end{document}